# Alleviating Bottlenecks for DNN Execution on GPUs via Opportunistic Computing

Xianwei Cheng
Department of Computer
Science and Engineering
University of North Texas
xianweicheng@my.unt.edu

Hui Zhao
Department of Computer
Science and Engineering
University of North Texas
hui.zhao@unt.edu

Mahmut Kandemir
Department of Computer
Science and Engineering
Pennsylvinia State University
kandemir@cse.psu.edu

Saraju Mohanty
Department of Computer
Science and Engineering
University of North Texas
saraju.mohanty@unt.edu

Beilei Jiang
Department of Computer
Science and Engineering
University of North Texas
beileijiang@my.unt.edu

*Abstract*—Edge computing and IoT applications are severely constrained by limited hardware resource. This makes memory-consuming DNN frameworks not applicable to edge computing. Simple algorithms such as direct convolution are finding their way in embedded machine learning. As one of the most widely used platforms for DNN acceleration, GPUs face the bottleneck of on-chip bandwidth. This work introduces a GPU DNN execution architecture that targets on relieving the on-chip bandwidth bottleneck by reducing data movement through opportunistic computing. We first investigate data access patterns in the hardware view rather than the software view. Then we propose two opportunistic computing techniques to predictably perform computation when data is available with the help of assistant warps. By moving computation to data, our techniques are able to significantly reduce data movement and relieve the DNN execution bottleneck. Our evaluation results show that the proposed technique can improve DNN application performance as much as 55%.

## I. Introduction

In recent years, deep learning techniques have seen great success in many application domains such as speech processing, computer vision, and natural language processing. Researchers have recently been investigating the role of accelerators as a solution, including GPUs, ASICs and FPGAs. ASICs and FPGAs usually have better performance and power efficiency than GPUs. However, they employ specialized design tailored for a specific type of DNN and need to modify software stack which restricts their application realms. On the contrary, GPUs are easier to program and require no specialized modification to the programming models. Therefore, GPUs are widely used commodity accelerators in the DNN execution and become the most widely used platform for DNN training [1].

It has been shown that a critical performance bottleneck in executing DNN applications on GPUs is the on-chip bandwidth [2]. In order to reduce data movement, prior research has proposed several techniques such as pruning and data reuse. However, these optimization techniques either work at the application level or are from the view of the software programmers but ignore the underlying hardware architecture. For example, some pruning techniques are based on the software-defined filters without considering how the filters are organized and stored in hardware memory [2]. Therefore, such schemes cannot lead to optimal performance improvement because the real hardware bottleneck is not addressed, which is caused by the mismatch between the algorithm and the underlying hardware architecture.

Recently, edge computing/IoT becomes a new research frontier, which is greatly enabled by moving machine learning to the network edge. Machine learning on the edge usually executes on embedded systems which are severely constrained by limited computing resource and memory space. Therefore, machine learning algorithms that require a large amount of memory (e.g. im2col in cuDNN) are not suitable for embedded systems [3]. Instead, simple algorithms such as direct convolution find their way in embedded machine learning and can achieve high performance if carefully designed [3]–[6].

In this work, we investigate techniques for embedded machine learning by reducing DNN on-chip data movement through data reuse at the architectural level. Instead of using the software-defined format of a basic compute unit (e.g., a filter), we investigate the data reuse with respect to cache blocks which is the actual unit of data movement in hardware. A GPU consists of several computing Streaming Multiprocessors (SMs or shader cores). In a single SM, we found that computation between data located in a pair of L1 cache blocks usually follows some regular access patterns. It is highly likely that a data cache block is revisited many times because the filters it holds are involved in multiple computations. We also observed that a large amount of duplicated data exists among L1 caches in different SMs, so the data missing from one SM can be found in other SMs with high probability. Moreover, computation results for a DNN layer are not likely to be reused until after the accumulation operation at the end of the current layer and this results in low data dependency.

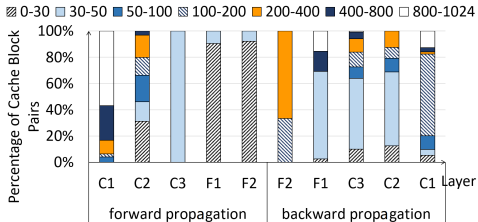

Fig. 1. Distribution of the number of computations between two cache blocks (LeNet5).

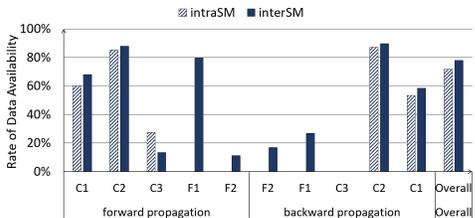

Fig. 2. Data availability when a GPU SM stalls due to load misses (LeNet5).

This motivated us to design an opportunistic computing mechanism, which takes advantage of data locality both within and across SMs to reduce data movement. Our scheme leverages near data computing by performing predicted computation **when** and **where** the data is available. The key idea is that there are repeated accesses to the same pair of cache blocks, separated in time. To the best of our knowledge, we are the first to propose such ***opportunistic computing architectures*** for DNN execution on GPUs. The execution here includes both training and inference. Our contribution is as follows: (1) Intra-SM Opportunistic Near Data Computing to predicatively execute computations *when the data is available* within an SM's L1 cache. (2) Inter-SM Opportunistic Near Data Computing which enables us to perform near data computing *where the data is available*. (3) Leverage of assistant warps to facilitate opportunistic computing.

## II. MOTIVATION

Computing convolution layers found in DNNs usually employs three methods: direct convolution [8], [9], unrolling-based convolution [10]–[12], and FFT(Fast Fourier Transformation)-based convolution [13], [14]. Direct convolution is the traditional way to compute convolution. A small window slides within an input feature map and a dot production between the filter bank and local patch of the input feature map is computed. Cuda-convnet2 [8] and Theano-legacy [9] are representative implementations of direct convolution. Unrolling-based convolution is a more efficient method by reshaping and duplicating the input and the filter bank to double large matrices using algorithms such as *im2col*. Then the final convolution is converted into a matrix-matrix production (GEMM). Typical implementations include Caffe [10], Torch-cunn [11], Theano-CorrMM [14], and cuDNN [12]. The FFT-based convolution first converts inputs and filter banks from the spatial domain to the Fourier domain, then those transformed matrices are multiplied in the Fourier domain and finally inverses back to the spatial domain [13], [14].

While unrolling and FFT based convolutions achieve good performance, in order to utilize the matrix-matrix multiplication routine, they require a large amount of memory space to store intermediate data [3]–[5]. As such, these methods are not suitable for the environment with limited resources, such as edge computing. Edge computing moves DNN execution to the resource-constrained devices in network edges such as embedded systems and mobile systems. Such systems cannot support GEMM or FFT computation even though they achieve better performance. Direct convolution, on the contrary, is well suitable for edge computing and embedded machine learning, due to its low memory requirement. However, new techniques need to be developed to enhance the performance of direct convolution. There have been several software-based techniques proposed [3]–[6] but few hardware solutions have been provided.

When GPUs allocate memory for input and weight data, they are allocated to different memory chunks. As a result, two pieces of data involved in a computation are from separate cache blocks. In this work, we call them *a computing cache block pair*. Ideally, if we can finish all computations when a computing cache block pair are both available in an SM's L1 cache, we can avoid stalls caused by load misses. However, in reality, it is not possible to schedule all computations between one computing cache block pair to one SM consecutively due to the high data parallelism in GPUs. The second reason is we need to balance the workload among SMs. Even if we know a computing cache block pair needs to perform a lot of computations, we still need to split the computations into different warps/threads and assign them to different SMs for better parallelism.

We characterized the frequency of computations between all computing cache block pairs in a real CNN application (LeNet5 [1]) in Figure 1. As can be observed, for the first convolution layer (C1) in the forward pass, about 57% of all the computing cache block pairs have more than 800 total computations between data from them. About 96% of the computing cache block pairs have computations more than 100 times for this layer. Similarly, we can observe a large amount of computation among computing cache block pairs for most of the layers except C3, F1 and F2. This means there are a lot of repeated accesses to many computing cache block pairs. We also characterize the possibility of the missing data to be found in other SMs in Figure 2: as can be observed, missing data can be found in other SMs with a probability as high as 86% for some layers and the overall probability of the whole application is 78%. To take advantage of the data locality, we propose opportunistic near data computing techniques detailed in the next section.

## III. PROPOSED DESIGN

### A. Intra-SM Opportunistic Near Data Computing

In this scheme, we propose to predict future computations based on patterns observed in data access history. Then we perform the computation when both the data blocks are available in the L1 cache of an SM. When the predicted computation is really scheduled to the SM, we can just use the predicted results directly without incurring SM stalls and data movement. We call this technique intra-SM opportunistic near data computing because it only takes advantage of intra-

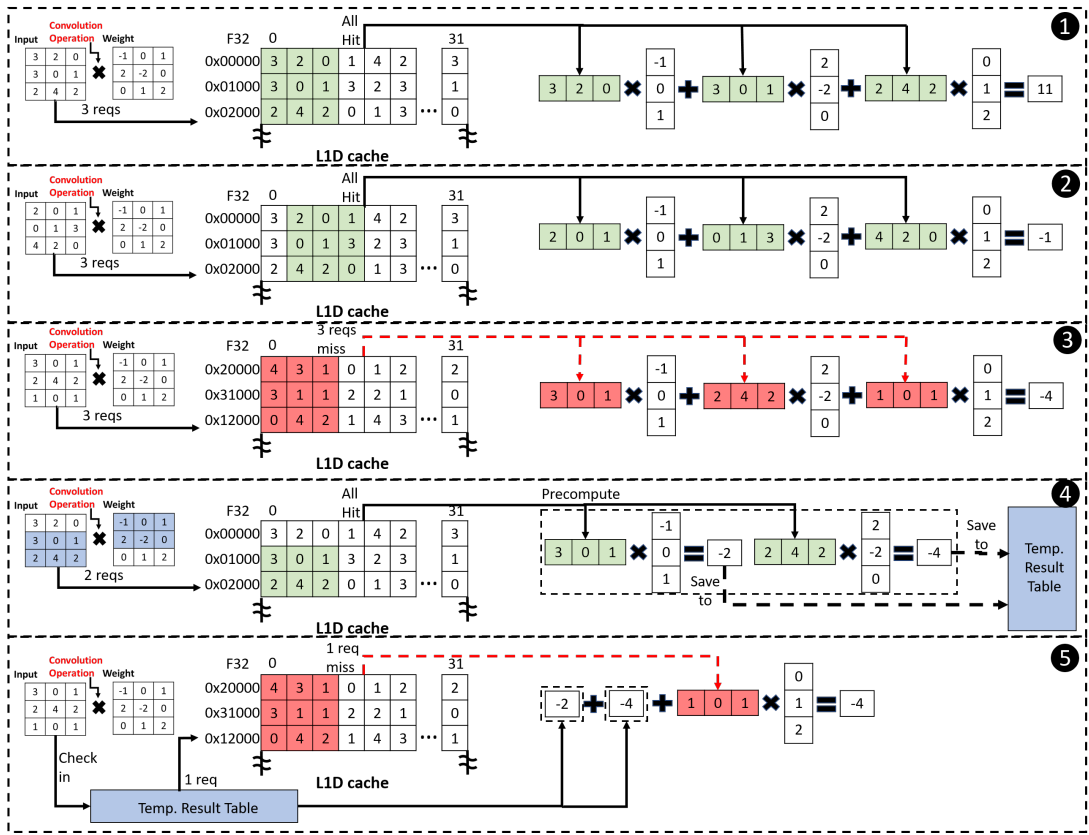

Fig. 3. An example of intra-SM opportunistic computing. The left side of each step shows the software view of data in the computation: input and weight windows; the middle shows data organization in the L1 cache; the right side shows convolution computation implemented in the step.

SM data locality and performs predicted computation when the data is available.

Figure 3 illustrates how our intra-SM scheme works using an example. To simplify the case, we only show input data stored in the cache. Because input data is allocated as a big chunk of memory, each cache block can only hold part of the data from one row of the input data and an input row takes up multiple cache blocks. In this example, we assume each cache block can hold 32 words and an input row has 256 words. Due to the GPU memory allocation policy, a 3x3 input window needs to access data from three cache blocks. As shown in ❶, the 3x3 input window on the left needs to access data from cache blocks with address (0x00000, 0x01000, 0x02000). We assume in ❶ that all three needed cache blocks are already fetched into the L1, so all three cache accesses are hits and the result can be calculated without stall. Then the input sliding window moves to the right by one stride in ❷. Since the data is already fetched by the last step, again all three cache accesses are hits. Then the procedure continues as the sliding window keeps moving to the right until reaches the end of the input row and this is not shown in the figure. Then in ❸, the input sliding window shifts down one row and restarts from the left end. Even though the new input window has its first two vectors overlapping with ❶, but because the cache blocks are already replaced during earlier steps with new data (the red square in the cache), all three accesses result in misses. As a result, this warp has to stall to load data.

In our scheme, we execute the action illustrated in ❹ right after step ❶ as opportunistic computing. Because we predict that the two bottom vectors (in green color) will be used in future computations, we predictably execute the calculation and save the results in the Precompute Table. Then in our ❺ which corresponds to baseline ❸, we only need to wait for one missing data to finish the whole convolution by using the predicted results. It needs to be noted that, in this example, we slide the windows horizontally. If the windows slide vertically, we still have a similar issue of revisiting a cache block after the window makes a turn.

We added hardware to manage the intra-SM opportunistic computing called Precompute Table. We use this table to save predicted computations and their results. When an instruction is decoded, we extract the cache block address information of both operands. Then we use the address pair to search the valid entries of the Precompute Table. If the pair of addresses is found and the completion bit is set, we know that this computation has been finished and we can use the result. If we cannot find the address pair, we know the computation is not precomputed and continue with normal execution. If we found the address entry but the complete bit is 0, this means the computation is predicted but not able to finish yet, so we invalidate the entry and continue with normal execution. When this SM stalls due to no warp can proceed, we start executing the assistant warps created with information in the Precompute Table.

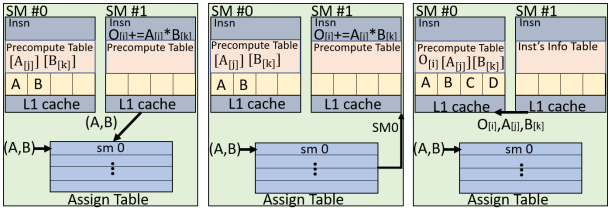

Fig. 4. Inter-SM Opportunistic near data predicted computing.

When the assistant warp finishes the computation, it will save the results into the Precompute Table. If at this time the normal warps are ready to execute, the control will jump back to normal warps. Otherwise, another assistant warp will be created to perform a new predicted execution. Because an SM stall often involves data movement from the L2 cache, it usually takes dozens of cycles. The assistant warp only takes a few cycles to perform the computation. Therefore, multiple assistant warps can be executed during one SM stall. To avoid wrongly predicted computation to saturate the table space, we periodically eliminate the oldest entries.

Our prediction method is based on the observation that the order of computations among input/weight maps exhibit regular patterns. As shown in the example in Figure 3, the sliding windows have several overlaps in data when sliding horizontally and vertically. Sliding horizontally can take good advantage of data locality in cache blocks. Our technique aims to utilize data locality when a sliding window moves down to a new row. This can be achieved by performing precomputation before we have all data in a sliding window available, as illustrated in Figure 3. In other words, we propose to use part of the data in the current input window to perform partial convolution with vectors in the weight window. An important step is to know what vectors will perform future convolution and this is not hard to predict, due to the regular movement of the windows. Algorithm 1 shows the detail of our prediction logic. Based on the window size and current input and weight vector address, we generate vector addresses for predicted computation. This is achieved by moving the input vector down and find all possible future computations with weight vectors until we reach the bottom of the window. Still using Figure 3 as an example, at the step 1, we can make three predicted computations: by moving the window down by one row, we get [3,0,1] x [-1,0,1] and [2,4,2] x [2,-2,0]; by moving the window down by two rows, we get [2,4,2] x [-1,0,1].

### B. Inter-SM Opportunistic Near Data Computing

Intra-SM opportunistic computing takes advantage of per SM data locality and performs future computations when data is available. If the prediction is not successful, an SM still needs to fetch the data from the L2 cache. Since each SM still keeps a copy of the data and is ignorant of other SMs' data locality, there still exists a lot of duplicate data in SMs and leads to redundant data movement. If we can share the data among the SMs, then we can not only reduce data movement but also improve each SM's L1 cache effective capacity. This motivates us to propose a second technique called inter-SM opportunistic near data computing. We organize all the SMs into several clusters (similar to the AMD's GPU cluster) and we move computation to an SM that has the data located in its L1 cache within each cluster.

Figure 4 shows our cluster-based inter-SM opportunistic near data computing architecture. Inside each cluster, we build a lookup table to manage the mapping of a computation to an SM called Computation Assignment Table (Assign Table). Each entry of this table is indexed by the address of a computing cache block pair. The content of each entry is a id of an SM inside the cluster. For example, an entry indexed by (A, B) with a value of 3 means that currently SM 3 has both cache blocks and computation from other SMs can be moved to SM 3 for near data computing. When an SM performs computation and encounters a stall due to data miss, that SM first checks if the table has an existing entry indexed by the two cache blocks involved in the computation. If there is no such an entry, then the SM loads the missing cache block from the L2 cache and creates a new entry in the Assign Table. We update the table with changes in cache block replacements. If a cache block is replaced from an SM's L1, then we search the table and remove all entries related with this cache block address. To reduce area overhead and search latency, we limit the table size to be 512 entries.

For each SM, we perform not only assigned computations but also predict future computations. This is similar to the intra-SM scheme and the only difference is this time the computation is not from a thread scheduled to this SM but is instead moved from another SM. The SM that assigns its computation to another SM does not need to read the result back. Instead, it sends the output address to the assigned SM to finish the write back to memory directly. This is due to the characteristics of the MAC operations in DNNs where the computation result of multiplication will only be used by an accumulator which is implemented as an atom add operation at a memory location.

Next, we describe the overall workflow of the inter-SM scheme which is shown in Figure 4. Two SMs 0 and 1 are in the same cluster. SM 1 needs to perform computation between two cache blocks A and B and encounters a cache miss. Then SM 1 searches the Assign Table and finds that SM 0 has both cache blocks. Then SM 1 sends its input and output addresses to SM 0. SM 0 searches its computation results in the Precompute Table to see if this computation has been performed predictably. If not, a new entry will be added to the Precompute Table so the next assistant warp will perform the computation. At the same time, SM 0 will skip this load stall and continue with its other operations. The inter-SM can improve performance in three ways: (1) we perform predicted computation during normal SM stall similar to the intra-SM scheme; (2) we can greatly reduce data movement and schedule computation to where the data is located; and (3) duplicated data is reduced among L1 caches inside a cluster which will improve cache hit rate.

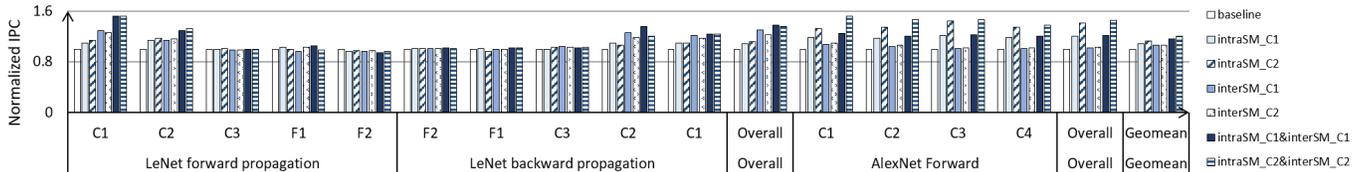
Fig. 5. Normalized Performance(IPC).

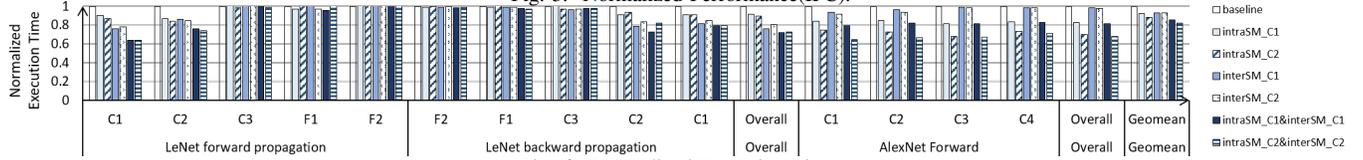
Fig. 6. Normalized Execution Time.

## C. Design of Assistant Warps

Our proposed schemes perform Near Data Computing by generating special warps which we call assistant warps to execute predicted computations during an SM's stall time. It has been shown that in GPUs, there is under-utilization of computing resources such as registers and on-chip shared memory. For example, there are many registers that are left unallocated to any thread block because the number of available registers is not a multiple of that required by a thread block. Such unused resources can be used by us to create assistant warps to alleviate the bottlenecks in GPU execution. To reduce overhead, we only allocate resource to one assistant warp every time. When an assistant warp finishes its execution, we reuse its resource to create a second assistant warp. Each time we create only one assistant warp and it will not encounter a stall because all its data is available. As a result, each assistant warp only takes a few cycles to finish. Since an SM stall usually takes tens of cycles (e.g. waiting for data from the L2 cache), we can execute multiple assistant warps one after another during this stall time.

Our primary motivation is to utilize existing idle resources for assistant warp execution. The proposed assistant warps is a pure hardware solution because their function is just to finish the predicted and assigned computation. There is no programmer or compiler involvement in the creation and execution of the assistant warps. Every assistant warp is a set of instructions issued into the SM pipelines. Just like regular instructions, assistant warp instructions are executed in lockstep across all the SIMT lanes. Our assistant warps own a separate context that is allocated in each SM (e.g., registers, local memory) but does not reduce the number of threads that can be scheduled on a single SM. Ideally, an assistant warp consumes resources and issue cycles that would otherwise be idle.

## IV. EVALUATION

We use GPGPU-Sim 3.2.2 [15] to simulate our proposed opportunistic near data computing architecture. Our baseline is a GeForce GTX 480 GPU using an 8x8 2D mesh to connect 56 SMs and 8 MCs. Table 1 shows the configuration used in our evaluation. For inter-SM scheme, we divide the SMs into 8 clusters. We implemented the CNN applications using direct convolution in a similar approach as in [16]. The applications we used in this work are LeNet5 and AlexNet [7]. We are able to run LeNet5 to the completion because the scale of this neural network is relatively small. However, we were not able to finish the whole application of AlexNet on GPGPU-Sim due to the prohibitively long simulation time. Instead, we selectively run the first 4 convolution layers of AlexNet because convolution layers are the most computation-intensive layers and most significantly affect the overall performance.

TABLE I
SYSTEM CONFIGURATION.

| SM | 56 SMs, 1.4GHz, SIMT width=8 |
|---|---|
| Warp Scheduler | Greedy-Then-Oldest |
| Shared Memory | 48 KB |
| Cache | 2KB L1 I-Cache (4 sets/4 ways LRU), 16KB L1 D-Cache (32 sets/4 ways LRU), 64KB L2 Cache per MC (8 way LRU) |
| Memory Model | 8 MCs, 924 MHz |
| NoC | 128-bit channel width, 2-stage pipeline, 16-byte flits, 1-cycle link latency, X-Y routing, vc buffer depth=4 |

To evaluate the tradeoff between resource overhead and performance, we experimented with two configurations in our schemes: intraSM_C1 and interSM_C1 are low cost designs with 256 entries in the Precompute table and 512 entries in the Assign Table. The table sizes are doubled to 512 and 1024 in intraSM_C2 and interSM_C2 respectively. The performance evaluation of our proposed schemes is shown in Figure 5 and Figure 6. As can be observed, our Intra-SM Opportunistic Near Data Computing contributes to 8.0% (intraSM_C1) and 12.6% (intraSM_C2) performance improvement and 9.0% (intraSM_C1) and 13.0% (intraSM_C2) execution time reduction on average. Forward propagation C2 layer in LeNet has the highest performance gain and execution time reduce for both intraSM_C1 (14.2% and 13.4%) and intraSM_C2 (17.5% and 16.2%), whereas, in AlexNet, it is the forward propagation C3 layer that improves performance and reduces execution time the most with intraSM_C1 (21.1% and 18.2%) and intraSM_C2 (44.7% and 32%). For Inter-SM Opportunistic Near Data Computing, it improves the performance by 6.5% and reduces execution time by 6.7% for both interSM_C1 and interSM_C2 on average.

There is 15.0% performance gain and 14.2% execution time reduction on the average. When more resource is allocated, 20.2% performance improvement and 18.0% execution time reduction can be achieved in the second type of configurations. The most significant performance gain is from the forward C1 layer in LeNet which is around 55%. It is worth noting that, although there is no significant improvement for some unaffected layers, our technique does greatly reduce the execution time of most convolution layers. When we evaluate the overall performance with all layers included, LeNet achieves a performance gain of 40.0% and AlexNet improves by 43.1%.

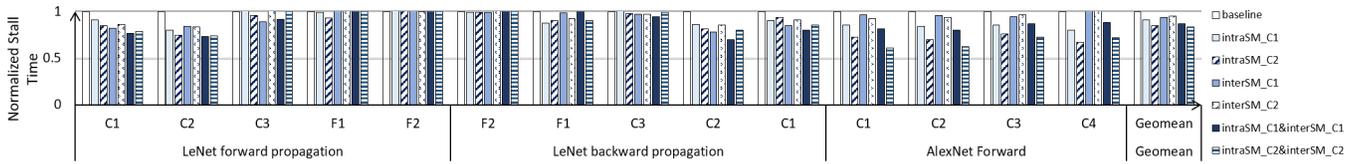

Fig. 7. Normalized Stall Time.

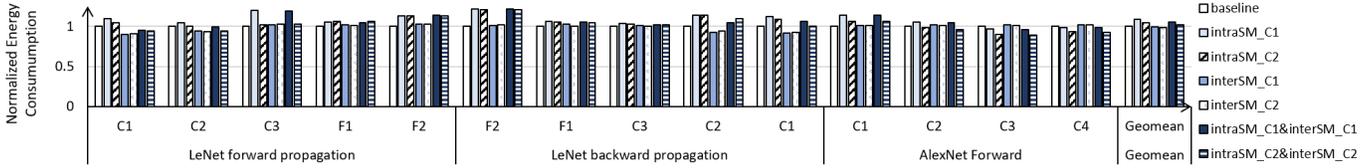

Fig. 8. Normalized Energy Consumption.

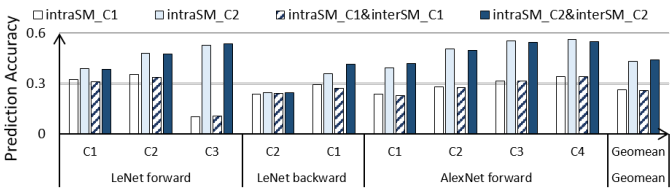

Fig. 9. Prediction Accuracy.

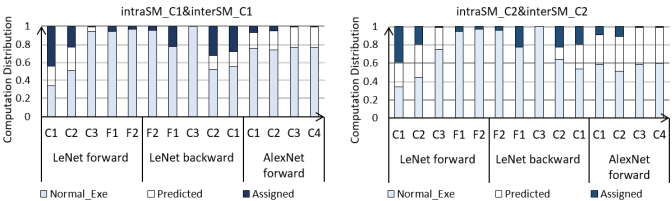

Fig. 10. Computation distribution.

We also evaluate the SM stall time in each technique and the result is shown in Figure 7. Compared with baseline, the intraSM_C1 can reduce stall time by 11% and intraSM_C2 can further reduce it by 15%. When the table sizes are doubled, the stall time reduction is more significant: 20% for LeNet and 26% for AlexNet. The Inter-SM schemes can reduce the stall time to 93.6% (interSM_C1) and 95.4% (interSM_C2) of the baseline averagely. When both intraSM and interSM techniques are combined, the average stall time can be reduced by 13% and 20% on the average across all workloads. We evaluate the energy consumption when employing our techniques and the results are shown in Figure 8. As can be observed, the inter-SM scheme brings nearly zero power overhead. This is because the energy consumed by Assign Table is offset by the energy saved through reduced NoC data movement and memory accesses. Energy consumed in Intra-SM scheme increases around 6% which is mostly caused by the static power of table buffers and mispredicted computations.

We also evaluated our prediction accuracy in Figure 9. As can be observed, the accuracy doubled when we increase the Precompute table size in intraSM_C2, from 26.2% to 43.3%. We can also observe that combined with the inter-SM scheme, a small degradation occurs. This is because the inter-SM scheme moves computation to other SMs which makes the prediction less accurate. Figure 10 shows the distribution for normal, predicted (intra-SM opportunistic computing) and assigned computations (inter-SM opportunistic computing). We can observe that the predicted and assigned computations occupy around 50% of total computations for forward C2, backward C2 and C1 layer, with more than 50% for the forward C1 layer.

Finally, we evaluate the overhead of our techniques. The proposed precompute table has 256 entries, with 64 index bits, 1 valid bit, 1 complete bit and a result of 32 bits. The overall size of one precompute table is around 3.1 KB per SM. For the Assign Table, there are 512 entries in each cluster of 7 SMs. For each entry, there are 50 block address bits, and 3 bits of assigned SM ID. Therefore each Assign Table is 512 * (50 + 3) = 3.4 KB per cluster. Each assistant warp needs 3*4*32=384B registers and we added 5.036KB in total per SM, compared to the overall storage of a baseline SM (118KB), our overhead is 4.27%.

## V. RELATED WORK AND CONCLUSION

Imani et.al explored the using of a resisted nearest content addressable memory blocks (NNCAM) for frequently accessing operations storing [17]. Du et.al employed inexact computing theory into neural network architectures for its sizable tolerance with errors [18]. Kim et.al proposed a kernel decomposition method used for binary weight neural networks for operation reduction [19]. Zheng et.al proposed a kernel transformation diagram for operation reduction that fits for both binary and ternary weight neural networks [20]. Our proposed technique is orthogonal to these work because we are providing architectural level solutions and our technique can be combined with other work. In this work, we propose techniques for DNN execution in edge computing/IoT where memory consuming algorithms are not applicable. Our techniques reduce data movement by taking advantage of opportunistic computing and can effectively relieve the execution bottleneck to improve system performance.